# Extraction of Core Contents from Web Pages

**Sandeep Sirsat**
Associate Professor and Head
Department of Computer Science
Shri Shivaji Science & Arts College, Chikhali, (Maharashtra)

**ABSTRACT: The information available on web pages mostly contains semi-structured text documents which are represented either in XML, or HTML, or XHTML format that lacks formatted document structure. The document does not discriminate between the text and the schema that represent the text. Also the amount of structure used to represent the text depends on the purpose and size of text document. No semantic is applied to semi-structured documents. This requires extracting core contents of text document to analyse words or sentences to generate useful knowledge. This paper discusses several techniques and approaches useful for extracting core content from semi-structured text documents and their merits and demerits.**

## I. Introduction:

Now a days, web is growing rapidly with huge amount of information is available in heterogeneous formats, such as, web pages, web archives, news wires, technical documents etc. Extracting high quality content efficiently from these web pages is crucial for many web applications such as, information retrieval, information extraction, topic tracking, text categorization and summarization. Many researchers have studied the problem of extracting content from web by means of different scientific tools in a broad range of application domain. These techniques deals to locate the specific web news page by interacting with web sources and extract the content stored in it. For example, if the source is an HTML web page, the extracted information consists of elements in the page as well as the full-text of the page itself. This requires pre-processing the extracted content, discovering the knowledge by converting it into a convenient structured form and storing it for later usage.

Most IE systems use wrapper generated by wrapper induction system for content extraction from Web page. For an information source, only one wrapper is generated. Wrapper generally provides single uniform query interface to access multiple information sources. It wraps information source using information integration system and access information source without changing its core query answering mechanism.

Different approaches use some techniques of Web mining, such as classification and clustering, to extract content from Web news page.

Some approaches use statistics to extract content from Web page. These approaches usually perform

Web page extraction techniques based on the method of web page modelling are classified into the following two main categories, which utilizes the features of Document Object Model (DOM) structure.

i. tag sequence based, and

ii. tree based

## II. Problems in Web IE:

Wrappers that are used to collect information from the Web server needs to collect the resulting pages via HTTP protocols, perform information extraction to extract the contents in the HTML documents [1]. Most wrappers are template dependent and usually generated for only one information source [2][3]. This increases the cost of maintenance of wrappers for thousands of web sites. The task of Web IE for collecting the information from such heterogeneous sources requires developing of wrapper induction (WI) systems that varies in scale depending on the text type, domain, and scenario [1].

Approaches that use techniques of classification and clustering improve the accuracy of information extraction. However, these approaches have limited ability for scalable extraction, because of human interventions and the complexity of the underlying algorithms. Some Web IE systems extract content from Web pages which use approaches based on machine learning (ML) techniques [4], or Natural Language Processing Techniques (NLP) [1][5]. The machine learning techniques can be used across sites, but again need to be re-trained as sites evolve.

Approaches based on statistics need to determine weights or thresholds using empirical experiments. Difficulty arises to find one set of weights or thresholds to satisfy all news pages coming from various heterogeneous news sources.

## III. Tag sequence based

Tag sequence based techniques view web page as a long sequence consisting of HTML tags and text fragments. The basic idea of traditional pattern reduction techniques is applied in order to find a template. Most of these techniques include substantial features of the DOM tree. Here we present the following three template/site independent novel approaches for IE:

### a) V-Wrapper

Shuyi Zeng and et al proposed a novel template independent news extraction approach to identify the core content of news articles based on visual consistency [3]. The intuitive presumption imposed on identifying the main content block of news page includes the following visual features:-

i. The area that specifies contents is relatively larger than other page objects around it.

ii. Usually, a bold-faced line at the top of the block is the news headline.

iii. It is mostly occupied by contiguous text paragraphs, sometimes mixed with one or two illustration pictures.

iv. The centre of block is close to that of the whole page.

This approach uses MSHTML library to render, and parse HTML source to obtain a visual tree, which provides interfaces to access information from any DOM nodes. The basic visual features include: position features, size features, rich format features, and statistical features. This approach uses parent-child relation between two blocks, which defines a set of extended visual features between a child block and its parent block. This parent-child relationship is used to match any two blocks as to paternity, inspite of their nested depth. This effectively handles the topological diversity of DOM trees. The V-wrapper is trained to learn the human behaviour involved in browsing news pages by using a set of manually labelled pages. This process of learning behaviour is a two-step process.

i. Locating Features: V-wrapper first, locate roughly the main block based on all kinds of visual features presented in the page and excluding all irrelevant content like advertisements, banners, etc.

ii. Identifying Features: It looks into the main block more carefully and identify the headline, the news body, and so on.

These two steps normally require different visual features, which have been discussed above.

Extraction algorithm using generated V-wrapper also has two steps.

i. To find and extract leaf blocks whose parents are positive inner blocks as candidates for target information.

ii. Labelling different types of information from candidate blocks obtained in the first step. This involves using the leaf block classifier to match each candidate block with the labels. The output of this step is a post label L(p) for test page p. It indicates the target information, which must be extracted from p. The recursive algorithm using top-down approach is given below [3].

Algorithm to Identify Candidate Blocks (P)

```
begin
Parse p to visual block set B;
pageBlock := page level block in B;
plBlocks := empty block set;
Extract(pageBlock, plBlocks) ;
return plBlocks;
end
```

Algorithm for Extracting (b, plBlocks).

```
begin
if b is a inner block then
        if b is labeled as PI then
        childBlock := the first child block of b;
        while childBlock is not NULL
        Extract(childBlock, plBlocks);
        childBlock := childBlock's next sibling;
        endwhile;
        endif;
        else
        Add b to plBlocks;
endif;
end
```

### b) CoreEx

J. Prasad and et al. [4] developed a simple heuristic technique using a tag-base approach, called CoreEx to automatically extract the main article from online news. CoreEx uses a Document Object Model (DOM) tree representation of each page, where every node in the tree represents an HTML node in the page. This technique is based on analyzing the text and number of links in every node, and uses a statistical measure to determine the node (or a set of nodes) most likely to contain the main content.The simple heuristics considered in DOM node or set of nodes contains significantly more amount of text than the links. A Java Library HTML parser is used to obtain its DOM tree, which counts the amount of text, and links to recursively score every node in the DOM tree as shown in figure 1.

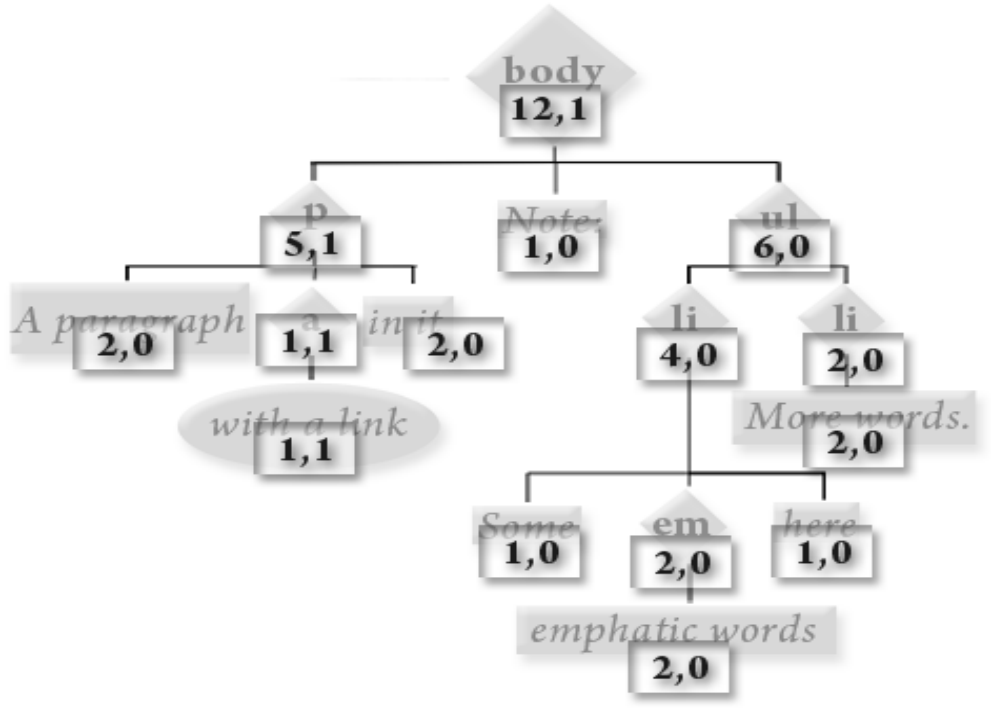

**Figure 1: DOM tree annotated with the word and link counts.**

It maintains two counts: *textCnt* and *linkCnt*, to obtain score for every node in DOM tree. *textCnt* holds the number of words contained in the node or it includes the sum of the words in the subtree for interior nodes. Similarly, *linkCnt* holds the number of links in or below the node. It then computes the node score using a weight scoring function as:

$$weight_{ratio} \times \frac{textCnt - linkCnt}{textCnt} + weight_{text} \times \frac{textCnt}{page_{text}} \quad \text{.........................................(1)}$$

Where $weight_{ratio}$ and $weight_{text}$ are the weights assigned to both the components proved to have performed well for the chosen values 0.99 and 0.01, respectively [17]. $page_{text}$ is the total text that the webpage contains. The weighted score function shown above is not feasible for extracting content distributed across several nodes. Hence J. Prasad and et al have suggested modified algorithm for non terminal node given in following algorithm

Algorithm for non-terminal nodes N

- Initialize
  textCnt(N) = 0 and linkCnt(N) = 0
  set S(N) as an empty set.
  setT extCnt(N) = 0 and setLinkCnt(N) = 0
- For every child of N,
  textCnt(N) = textCnt(N) + textCnt(child)
  linkCnt(N) = linkCnt(N) + linkCnt(child)
- Calculate
  childRatio = $\frac{textCnt - linkCnt}{textCnt}$
  If childRatio > Threshold
  add the child to S(N).
  setT extCnt(N) = setT extCnt(N)+textCnt(child)
  setLinkCnt(N) = setLinkCnt(N)+linkCnt(child)
  Store S(N), textCnt(N), linkCnt(N), setT extCnt(N)
  and setLinkCnt(N)

The nodes are now scored based on their setTextCnt and setLinkCnt. For the node with the highest score, return the corresponding set S as the set of nodes that contains the content.

CoreEx automatically convert XHTML page into plain text without expensive. J. Prasad observed that the performance is slightly below the baseline of prior work in terms of precision and recall used in IE process.

### C. ECON

Yan Guo and et al presented another approach named ECON to fully extract the entire content from web news pages automatically [2]. ECON uses HTML parser to convert the news Web page into a DOM tree and utilizes substantial features of the DOM tree. It presumes a key observation that, actual content of news Web page contains much more punctuation marks than noise in the same news Web page. ECON uses this key observation with some other to find a snippet-node to wrap the part of the content of news, and then backtracks from the snippet-node until a summary-node is found. It then extracts the actual content of news by removing noise from the subtrees of the summary node.

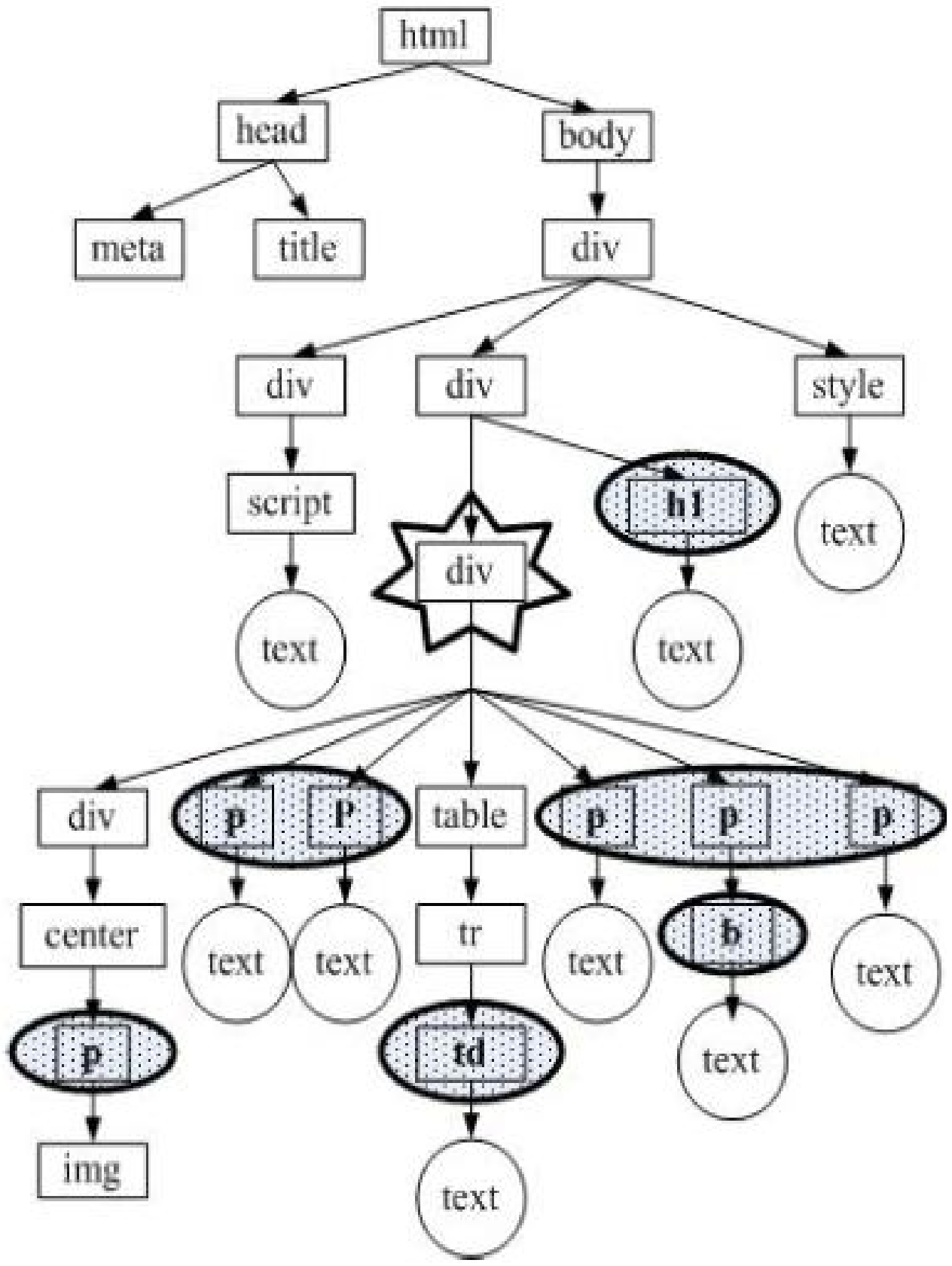

**Figure 2: A DOM tree of a web news page.**

Algorithms of ECON [2] are as follows:

#### i) Algorithm of Joint-para

It has been observed from a DOM tree of a Web news page that sometimes the entire text of news is broken into many short pieces by some nodes such as <p> and <br>. If there is a long piece of noise, there is a possibility of wrongly considering the piece of noise as the start point of backtracking. The algorithm of Joint-para will merge short pieces of text that helps in finding correct start point of backtracking. While merging, Joint-para tries to prune some noisy nodes embedded within some of the subtrees. The Joint-para algorithm works as below.

i. It first get a big-node as an input.
ii. Checks its brother nodes to find a text-node-set.
iii. Gets the text-para of the text-node-set and compute the punc-num of the text-para.
iv. Checks If punc-num = 0, the text-para will be regarded as noise and will not be output. Meanwhile, all the nodes that together wrap the noise piece are pruned.
If the punc-num not = 0, the text-para will be output.

### ii) Algorithm of Extract-news

The heuristics to detect when to stop backtracking is based on observation: While backtracking from node1 to node2, if the content of news wrapped by node2 is more than node1, node1 must not be the summary-node, and the node-puncnum of node2 must be more than node1. If node1 is the summary-node, then there will be the following two cases:

i. The information wrapped by node2 is equal to node1, so the node-punc-num of node2 must be equal to node1;
ii. There is more noise wrapped in node2 than node1, and the extra noise does not contain any period and comma, so the node-punc-num of node2 must be equal to node1.

Algorithm for Extract-news is as given below.

i. Input news web page, parse them to DOM tree, transverse the DOM tree and perform the algorithm of Joint-para for each big-node to get all text-paras.
ii. Select one node randomly from the text-nodeset that wraps the longest text-para and considers it as one snippet-node, and starts backtracking from the snippet-node. When backtracking from node1 to node2, it calculates the node-punc-num of node1 and that of node2, respectively.
iii. Compute the distance calculating the difference as: Distance = node-punc-num(node2) - node-punc-um(node1).
iv. Stop the process of backtracking on the following condition:
If distance = 0 for the first time, consider the child-node as the summary-node.
v. Extract the content wrapped by the summary-node as the entire content of news.

The definitions used in this algorithm are:

**Big-node** - means a node having tag-name either <p> or <br> or <h1> or <h2> or <h3> or <h4> or <h5> or <h6> or <strong> or <em> or <br> or <b> or <i> or <tt> or <font>, and not <script> or <style>.

**Punc-num** is a number of periods and commas appear in a text para, and the punc-num in the text wrapped by a node is called as **node-punc-num.**

## IV. Tree based Techniques

Web pages are naturally semi-structured in nature and represented as a labelled ordered routed tree called DOM tree. The DOM tree has been successfully exploited for the purpose of web content extraction in a number of techniques. These techniques mostly rely on analysing the structure of the target pages.

These techniques use the concept of tree edit distance to evaluate the structural similarity between pages [8][9]. The problem of computing the tree edit distance between trees is a variation on the theme of the classic string edit distance problem. Given two labelled ordered rooted tree A and B, the problem is to find a matching to transform tree A into tree B (or the vice-versa) with the minimum number of operations [7]. The set of possible operations performed on tree include node deletion, insertion or replacement. For each of this operation, a cost is applied and hence the task turns into discovering a mapping with minimum cost between the two trees (i.e., finding the sequence of operations of minimum cost to transform A into B).

The above presumption formally encoded in the definition of mapping is defined by Reis and et al [7].

*Definition: A mapping M between two trees A and B is defined as a set of ordered pairs (i, j), one from each tree, satisfying the following conditions for all* $(i_1, j_1)$ , $(i_2, j_2) \in M$.

- $i_1 = i_2$ , iff $j_1 = j_2$
- $T_1[i_1]$ *is on the left of* $T_1[i_2]$ *iff* $T_2[j_1]$ *is on the left of* $T_2[j_2]$
- $T_1[i_1]$ *is an ancestor of* $T_1[i_2]$ *iff* $T_2[j_1]$ *is an ancestor of* $T_2[j_2]$

The notation $T[i_x]$ - indicates the $x^{th}$ node of the tree T in a pre-ordered visit of the tree. This definition establishes a number of consequences as follows:

- Each node must not appear more than once in a mapping
- The order among siblings nodes is preserved
- The hierarchical relations between the nodes are unchanged.

A number of techniques based on this approach was proposed [7][10] which provides the support to all three types of operations on nodes (i.e., node deletion, insertion and replacement).

### a) Simple Tree Matching Algorithm

A simple tree-matching algorithm presented by S.M. Selkow provides computationally efficient solution for the problem of the tree edit distance matching [10]. For optimizing the performance, it imposes the restriction - node replacement operation is not allowed during the matching procedure. The pseudo-code of the simple tree matching algorithm is given in figure 3.4 which adopts the following notation.

i. d(n) represents the degree of a node n (i.e., the number of first-level children).
ii. T(i) is the $i^{th}$ subtree of the tree rooted at node T.

Algorithm for SimpleTreeMatching ($T_1$, $T_2$)

if $T_1$ has the same label of $T_2$

m ← d($T_1$)

n ← d($T_2$)
for i = 0 to m do

M[i][0] ← 0;
for j = 0 to n do

M[0][j] ← 0;

for all i such that 1 ≤ i ≤ m do

for all j such that 1 ≤ j ≤ n do

M[i][j] ← Max(M[i][j − 1], M[i − 1][j], M[i − 1][j − 1] + W[i][j]) where W[i][j] = SimpleTreeMatching($T_1$ (i – 1), $T_2$ (j – 1))

return M[m][n]+1

Else

return 0

**Restricted Top-Down Mapping (RTDM) Algorithm**

Reis and et al. proposed the domain specific Web Data Extraction approach based on a tree edit distance algorithm [7]. The algorithm relies on a different definition of mapping called Restricted Top-Down Mapping (RTDM). The RTDM algorithm is based on post-order traversal of trees. In RTDM insertion, removal and replacement, operations are restricted to the leaves of the trees. The restricted top-down edit distance between tree A and B is defined as the cost of the restricted top-down mapping between them. To find the restricted top-down mapping between trees A and B, the algorithm first computes the linear cost with respect to the number of vertices in the trees. After grouping the vertices in the trees in equivalent classes, Yang's algorithm [11] is applied to obtain minimum restricted top-down mapping between the trees.

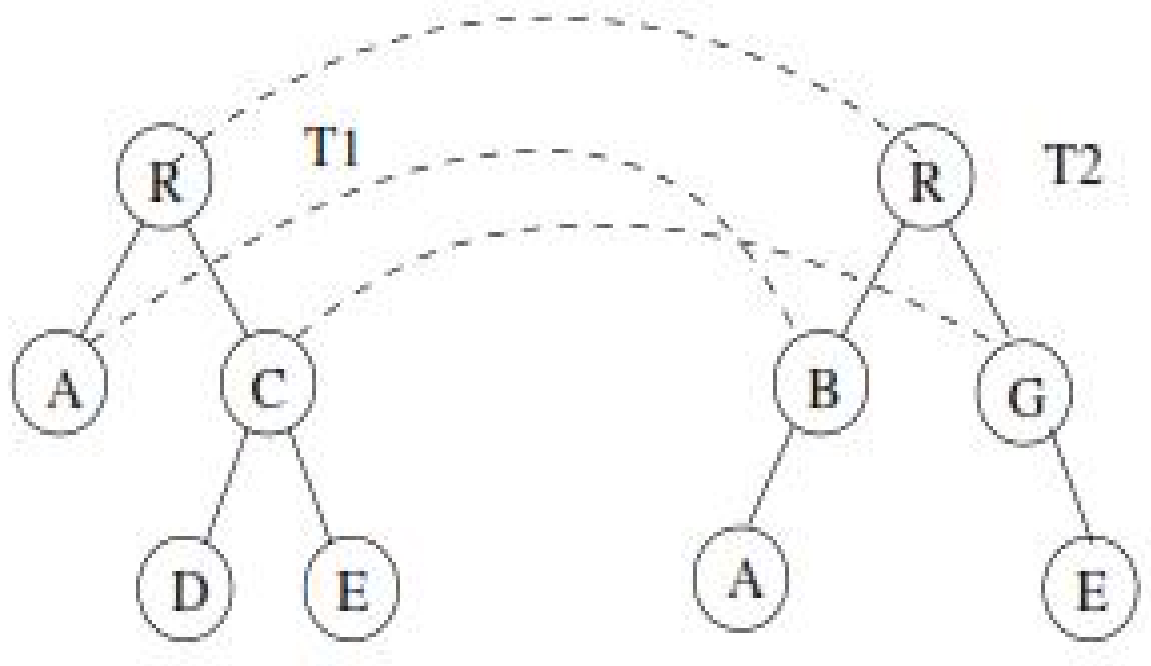

**Figure 3: A restricted top-down mapping example**

The pseudo-code of RTDM algorithm is [7].

The **RTDM** Algorithm

RTDM (*$T_1$* ,*$T_2$*, **$\epsilon$**: threshold)
begin
let *m* be the number of children of $T_1$ root
let *n* be the number of children of $T_2$ root
*M* [*i*, 0] ← 0 for all *i* = 0, ... , *m*
*M* [0, *j* ] ← 0 for all *i* = 0, ... , *n*
for *i* = 1 to *m*
for *j* = 1 to *n*
$C_i$ ← descendents($t_1[i]$)
$C_j$ ← descendents($t_2[j]$)
$d \leftarrow M[i-1, j] + \sum_k^{t_1[k] \in C_i} delete(t_1[k])$
$i \leftarrow M[i, j-1] + \sum_k^{t_2[k] \in C_j} insert(t_2[k])$
if $M[i-1, j-1] > \epsilon$
s ← ∞
elsif $t_1[i]$ and $t_2[j]$ are identical sub- trees
s ← 0
elsif
if $t_1[i]$ is a leaf
s ← $replace(t_1[i],\ t_2[j])$
s ← s + $\sum_k^{t_2[k] \in C_j} insert(t_2[k])$
elsif $t_2[j]$ is a leaf
s ← $replace(t_1[i],\ t_2[j])$
s ← s + $\sum_k^{t_1[k] \in C_i} delete(t_1[k])$
else
s ← RTDM ($t_1[i],\ t_2[j], \epsilon$)
fi
fi
*M* [*i, j*] ← *min(d, i, s)*;
end
end
return *M[m, n]*
end

## V. Conclusion:

Shuyi Zeng and et al demonstrated that the generated V-Wrapper has beneficial domain compatibility, and achieves extraction accuracy**.** However, V-wrapper requires a set of manually labelled news web pages for the purpose of training set. Most errors of V-Wrapper are caused by noise information (e.g., copyright information), which is visually similar to news text.

CoreEx is entirely automated but is requires prior conversion to XHTML, which is quite expensive. CoreEx is applicable to non-news genres as well. The scored DOM tree generated by CoreEx can differentiate the article pages and index pages of Web sites. CoreEx is not suitable to extract the contents of short news web page with accuracy. Its performance is below baseline in terms of precision and recall. This process does not include handling of news title and image captions that affects performance of the system.

Important aspect of ECON is that most of the features used are language-independent; and hence can be applied to news web page written in many languages such as Chinese, English, French, German, Italian, Japanese, Portuguese, Russian, Spanish, and Arabic. ECON can extract the contents efficiently and with high accuracy, but unable to deal with short news web pages with considerable accuracy in IE.

The mapping cost of the simple tree matching is O(nodes(A).nodes(B)), where nodes(T) is the function that returns the number of nodes in a tree T. The minimum cost of mapping ensures excellent performance when applied to HTML trees. The problem of mapping using tree edit distance is a difficult one. W. Chen and et al states that several algorithms, with different tradeoffs, have been proposed, but all formulations have some complexities [21]. Further, it has been proved by K. Zhang and et al that, if the trees are not ordered, the problem is NP-complete [22].

The RTDM algorithm can be applied to solve three important problems in automatic Web data extraction, namely: structure-based page classification, extractor generation, and data labeling. The RTDM algorithm has the worst case complexity of O(n1.n2). This occurs when the two trees being compared are all found identical, except for their leaves. But it performs much better due to the fact that it only deals with restricted top down mapping. This approach is intuitively based on the ambiguous assumption that the news site content could be divided into groups that share common format and layout characteristics. This is always not true for news web pages having heterogeneous structure and page layout.

**References:**

[1] Chia-Hui Chang, Mohammed Kayed, Moheb Ramzy Girgis, Khaled Shaalan - A Survey of Web Information Extraction Systems , IEEE TRANSACTIONS ON KNOWLEDGE AND DATA ENGINEERING, T Vol. 18, no.10,pp. 1411-1428, Oct. 2006.

[2] Yan Guo, Huifeng Tang , Linhai Song, Yu Wang, Guodong Ding, ECON: An Approach to Extract Content from Web News Page, 2010 12th International Asia-Pacific Web Conference, 978-0-7695-4012-2/ 2010 IEEE.

[3] Shuyi Zheng, Ruihua Song, Ji-Rong Wen, Template-Independent News Extraction Based on Visual Consistency, American Association for Artificial Intelligence (www.aaai.org), 2007.

[4] J. Prasad and A. Paepcke, "Coreex: content extraction from online news articles," in CIKM '08: Proceeding of the 17th ACM conference on Information and knowledge management. New York, NY, USA: ACM, 2008, pp. 1391–1392.

[5] Tanveer Siddiqui, U. S. Tiwari, Natural Language Processing and Information Retrieval, Oxford University Press Pages.

[6] Qiujun LAN- Extraction of News Content for Text Mining Based on Edit Distance Journal of Computational Information Systems 6:11 (2010) 3761-3777, November, 2010.

[7] Davi de Castro Reis, Paulo B. Golgher, Alberto H. F. Laender, Altigran S. da Silva, Automatic Web News Extraction Using Tree Edit Distance, ACM WWW2004, NewYork, USA, May17–22,2004.

[8] W. Chen. New algorithm for ordered tree-to-tree correction problem. *Journal of Algorithms*, 40:135–158, 2001.

[9] K. Zhang, R. Statman, and D. Shasha. - On the editing distance between unordered labeled trees. Information Processing Letters, 42(3):133–139, 1992.

[10] S. M. Selkow. The tree-to-tree editing problem. Information Processing Letters, 6:184–186, Dec. 1977.

[11] W. Yang. Identifying syntactic differences between two programs. Softw. Pract. Exper., 21(7):739–755, 1991.